\documentclass[twocolumn,jcp]{revtex4-1}
\usepackage[utf8x]{inputenc}
\usepackage{amsbsy,amssymb,amsmath,mathrsfs} 
\usepackage{graphicx} 
\usepackage{color}
\usepackage{listings} 
\usepackage[toc,page]{appendix}

\newcommand{\dif}{D}

\newcommand{\vect}[1]{\mathbf{#1}}

\lstMakeShortInline[columns=fixed]!

\begin{document}

\title{Solvation effects and contact angle saturation in electrowetting}

\date{\today}

\author{Nicolas Rivas$^{1,*}$}
\author{Jens Harting$^{1,2}$}

\affiliation{$^1$Helmholtz Institute Erlangen-N\"urnberg for Renewable Energy (IEK-11), Forschungszentrum J\"ulich, F\"urther Stra\ss e 248, 90429 N\"urnberg, Germany}

\affiliation{$^2$Department of Applied Physics, Eindhoven University of Technology, PO Box 513, 5600MB Eindhoven, The Netherlands}

\begin{abstract} 
Electrowetting of nanodrops is studied for aqueous electrolyte mixtures. We report a new method for controlling the degree of deformation and minimum attainable contact angle by varying the difference of solvation strengths between the two solvents, which determines the ratio of ion concentrations. This implies that the commonly observed saturation of the contact angle with the applied electric field can be suppressed by increasing the solvation energy difference, a finding of practical significance for micro and nanofluidics. Saturation is traced to be caused by the transfer of ions from the drop to the surrounding medium. Furthermore, we derive an expression, based on the electrokinetic equations, for the dependency of the drop's contact angle on the ion concentration and the strength of the solvation potential in the absence of external electric fields.
\end{abstract}

\maketitle





\textit{Introduction.---} The wetting properties of a drop in contact with a substrate change in an external electric field, a phenomenon referred to as electrowetting~\cite{mugele:2005,monroe2006electrowetting,zhao:2013,chen:2014}. While routinely used in technologies such as lab-on-a-chip systems and electronic displays, some basic behaviours of electrowetting still remain to be understood and/or properly controlled~\cite{chen:2014}. The degree of wetting, quantified by the contact angle, increases with the applied electric field, although only until a certain point, after which further increasing the strength of the electric field has little or no influence, an effect called contact angle saturation (CAS)~\cite{chevalliot:2012}. Although there exist many proposed mechanisms for CAS~\cite{mugele:2005,zhao:2013,chen:2014}, a general explanation is still lacking. Most probably there are a variety of factors involved, with their relative relevance a function of the typical length-scale, material properties and the use of either AC or DC voltages. The degree of uncertainty increases on the nanoscale, where experiments are harder to realize and several theoretical assumptions no longer hold. Most importantly, at the nanoscale the characteristic lengths of electrostatic interactions and charge distributions can be comparable to the drop size~\cite{liu:2012}, a limit of increasing interest which has received relatively little attention in the past. 

In this letter we investigate the effect of selective solvation on the wetting properties of nanoscopic sessile drops in electric fields. Selective solvation determines the partitioning of ion concentrations in the bulk of both solvents at equilibrium, and is quantified by the difference between the solvation chemical potentials of the solvents, i.e. the Gibbs transfer energy. Previous theoretical and numerical studies have so far disregarded the effect of the ions in the outer fluid even though, as we show here, the Gibbs transfer energy represents a novel control parameter of the wetting properties of drops as, for example, it can be tuned by modifying the concentration of water in aqueous electrolyte solutions~\cite{kalidas2000gibbs,marcus2007gibbs}. We derive an expression indicating a non-linear monotonous increase of the equilibrium contact angle with the Gibbs transfer energy, in good agreement with numerical solutions of the electrokinetic equations. With electric fields, we observe a transition from complete wetting to CAS as the Gibbs transfer energy is increased. CAS is found to be caused by the loss of ions from the drop, which eventually limits the electrostatic forces.


\begin{figure}
	\begin{center}
		\includegraphics[scale=1.0]{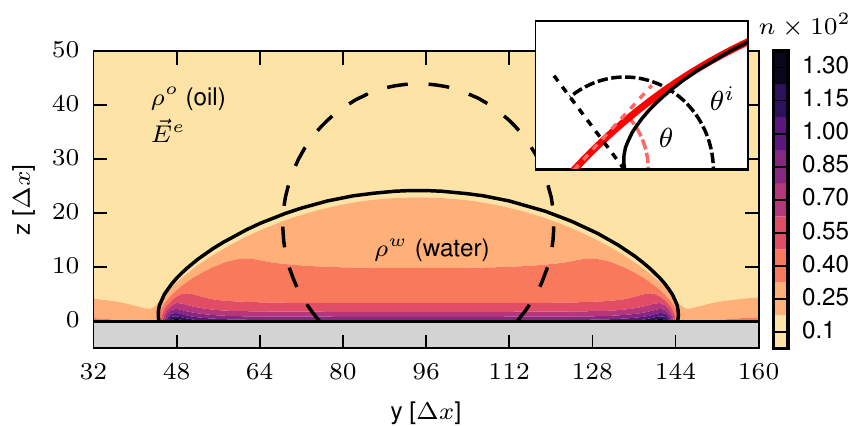}
	\end{center}
	\caption{(Color online) In black, contour line of the local concentration field $c = (\rho^a-\rho^b)/(\rho^b+\rho^a) = 0$, showing the drop deformation for different external field strengths $E_z^e = 0.01$ (dashed) and $E_z^e = 0.05$ (solid),  with $\Delta \tilde \mu = -4$, $\bar n = 0.001$. Colors correspond to the local ion concentration $n(\vect{r})$. The inset shows the region around the triple contact point (same data as main figure), and the definitions of the microscopic contact angle $\theta^i$, and the apparent contact angle $\theta$ defined by the extrapolation of a circular interface fit in red (gray).}
	\label{fig:drops}
\end{figure}

Our setup consists of an electrolyte suspension with a polar solvent $w$ (water) immersed in a less polar solvent $o$ (oil), as shown in Fig.~\ref{fig:drops}. Both fluids are conductive due to the flux of dissolved ions, with their respective conductivity given by the solvation potential, which sets the equilibrium ion concentrations in each fluid. Overall we wish to isolate the effects of solvation by considering the simplest scenario of electrowetting. Therefore we consider both fluids to have equal dielectric permittivities, and the substrate on which the drop rests to be apolar and uncharged. The height of the container is large enough so that the top boundary can be considered to be infinitely far away.  As the drops are expected to have azimuthal symmetry, two-dimensional systems are considered. The drop is deformed by an external electric field $\vect{E}^e$, applied perpendicular to the substrate in the $\hat z$-direction. 

\textit{Model.---} The solvents, the dissolved ions and the electrostatic interactions are described by means of the electrokinetic equations,
\begin{align}
	\label{eq:continuity}
	& \frac{\partial \rho^\sigma}{\partial t} 
	+ \nabla \cdot \left( \rho^\sigma \vect{u} \right) 
	= 
	0,
	\\ 
	\label{eq:navier_stokes}
	& \frac{\partial (\rho^\sigma  \vect{u})}{\partial t} 
	+ \nabla \cdot \left( \rho^\sigma \vect{u} \otimes \vect{u} \right)
	= 
	- \nabla \cdot (p \mathbb{I}) 
	+ \nabla \cdot \vect{s}
	+ \vect{F}^\sigma,
	\\
	\label{eq:nernst-planck}
	& \frac{\partial n^\pm}{\partial t} + \nabla \cdot (n^\pm \vect{u}\,)
	=
	\nabla \cdot \vect{j}^\pm,
	\\
	\label{eq:poisson}
	& \nabla^2 \phi^i
	=
	- \frac{q}{\varepsilon}.
\end{align}
The continuity and Navier-Stokes equations \eqref{eq:continuity} and \eqref{eq:navier_stokes} describe two (weakly) compressible fluids of density $\rho^\sigma(\vect{r})$ and hydrodynamic velocity $\vect{u}(\vect{r})$, with $\sigma \in (w,o)$. Here, $\otimes$ is the outer product, $p$ the pressure, $\mathbb{I}$ the identity matrix, and $\vect{s}$ the deviatoric stress tensor, $\vect{s} = \lambda (\nabla \cdot \vect{u}) \mathbb{I} + \eta (\nabla \vect{u} + \nabla \vect{u}^T)$, with $\eta$ the dynamic viscosity and $\lambda$ the bulk viscosity. The Nernst-Planck equation \eqref{eq:nernst-planck} describes the advection and diffusion of the concentration of two ionic species $n^\pm(\vect{r})$, cations ($+$) and anions ($-$), assumed to have unitary valences. The diffusive flux is given by Fick's diffusion plus a drift due to the forces applied to individual ions in the limit of low Reynolds numbers, $\vect{j}^\pm = \dif \nabla n^\pm + \mu \vect{f}^\pm$. The diffusivity $\dif$ is assumed to be homogeneous and equal for both types of ions. The mobility is given by Einstein's relation $\mu = D \beta$, where as usual $\beta \equiv 1/k_B T$, with the Boltzmann constant $k_B$ and the temperature $T$. Finally,  Poisson's equation \eqref{eq:poisson} relates the internal electric potential $\phi^i$ with the distribution of charge $q(\vect{r}) = e n^+(\vect{r}) - e n^-(\vect{r})$, where $e$ is the proton charge.

The force acting on the ions has two contributions, electrostatic and a pressure difference coming from chemical potential variations, $\vect{f}^\pm = q^\pm \nabla \phi + n^\pm \nabla \mu^\pm_s$. The total electric potential is the sum of an internal and external contribution $\phi = \phi^i + \phi^e$, with the corresponding fields $\vect{E} = \vect{E}^i + \vect{E}^e$. 

The solvation chemical potential $\mu^\pm_\text{s}$ is taken to depend linearly on the water density $\rho^w$~\cite{yabunaka2017electric},
\begin{equation}
	\label{eq:solvation_potential}
	\mu^\pm_\text{s}(\rho^w)
    = 
    - \frac{g^\pm}{\beta} \rho^w(\vect{r}),
\end{equation}
with $g^\pm$ a proportionality constant, which we take to be equal for cations and anions. This is just a first order model of interactions that are known to be more complex, although it is expected to hold at low ion concentrations~\cite{onuki2011phase}. Equal coupling constants $g^+=g^-$ imply that without an external electric field the system remains electrically neutral. In neutral systems at equilibrium the bulk ionic concentration ratio $\nu \equiv n_w/n_o = \exp(g\Delta \rho^w)$, where $n = n^+ + n^-$. 
The coupling $g$ is determined from experimental measurements of the Gibbs transfer energy $\Delta \mu^\pm = \mu^{\pm}_\text{s}(\rho^w_w) - \mu^\pm_\text{s}(\rho^w_o)$, with $\rho^w_w$ and $\rho^w_o$ the densities of water at the bulk of the water and oil phases, respectively. In our case $\Delta \tilde \mu \equiv \beta \Delta \mu = - g \Delta \rho^w \in (-4, 0)$, covering the range of many different ion types going from water to the most common aqueous organic solvents~\cite{kalidas2000gibbs,marcus2007gibbs}.

The external force term $\vect{F}^\sigma$ in Eq.~\eqref{eq:navier_stokes} comes from ion--solvent and solvent--solvent interactions, $\vect{F}^\sigma = \vect{F}^e + \vect{F}^i$. The former has in itself two more contributions, the reaction force of the friction coupling between ions and fluids $- \sum \vect{j}^\pm / \beta D$, and a solvation interaction force $\mu_s \nabla n$. This results in
\begin{equation}
	\label{eq:electric_fluid_force}
	\vect{F}^e = 
			   - \frac{\nabla n}{\beta}
			   - \nabla
				 \left(
			   		n \mu_\text{s}
				 \right)
			   + q \vect{E}.
\end{equation}
In what follows we refer to the individual terms in Eq.~\eqref{eq:electric_fluid_force} as, from left to right, osmotic, solvation and electric force. As the length-scales of the systems considered are much smaller than the capillary length, gravity is disregarded.

The solvent dynamics are resolved via the lattice Boltzmann method, which is based on the discrete Boltzmann equation
\begin{equation}
	f_d^\sigma(\vect{r}_i+\vect{c}_d\Delta t, t+\Delta t) - f_d^\sigma(\vect{r}_i,t) = \Omega.
\end{equation}
Here $f_d^\sigma$ is the single particle distribution function for solvent $\sigma$ in direction $d$, $\Omega$ the usual BGK collisional operator, $\vect{r}_i$ the coordinates of a regular cubic lattice with lattice constant $\Delta x$, and $\vect{c}_d$ the base vector in direction $d$ of a D3Q19 lattice~\cite{benzi1992lattice}. Even though we consider the dynamics in a plane, we simulate 3D systems in quasi-2D geometries, in order to have consistent results with possible full 3D geometries of future studies.

Fluid interactions, which determine $\vect{F}^i$, are modelled using the method of Shan and Chen~\cite{shan1993lattice}, tuned such that the water--oil surface tension $\gamma^{ow} = 0.03 m \Delta t^{-2}$, with $m$ the mass scale. This is a diffuse interface model which implies that the fluid--fluid interface has a finite width $\delta$. In our case $\delta \approx5\Delta x$, decreasing less than $5\%$ with increasing salt concentration.

The advection--diffusion equation \eqref{eq:nernst-planck} is solved via a finite difference discretization, and Poisson's equation \eqref{eq:poisson} is solved using a Fourier method. For further discussion of the model and details of the numerical methodology we point the reader to Ref.~\cite{rivas2018mesoscopic}. 


Special attention has to be paid to the solvent wetting model. We use a local interaction potential analogous to the solvent interaction model, following the methodology presented in Refs.~\cite{harting:2006,huang:2007,schmieschek2011contact}. The interaction parameter of fluids and solid is varied such that the equilibrium contact angle of the drop $\theta_0 \in (90^\circ,140^\circ)$~\cite{harting:2006}. This is the the apparent or macroscopic contact angle $\theta$, sustained by the substrate and the projection of the fluid/fluid interface away from the substrate, as shown in Fig.~\ref{fig:drops}. It was verified to follow Young's relation $\cos(\theta) = (\gamma_{os}-\gamma_{ws})/\gamma_{ow}$ in systems with no ions, with the surface tensions measured in a planar interface as the difference of the normal components of the stress tensor. This is the angle most commonly measured in experiments, due to the difficulties of accessing the region around the triple contact point (TCP) and measure the microscopic contact angle $\theta^i$ (see Fig.~\ref{fig:drops})~\cite{liu:2012}. We have also computed $\theta^i$ and observed similar qualitative behaviour although much smaller variations. Experimental measurements and theoretical derivations have suggested $\theta^i$ to be invariant on the applied electric field~\cite{mugele2007equilibrium,gupta:2010}, but we expect to see finite variations in our simulations due to the limited resolution at the TCP. 

In the following we relate the variations of $\theta$ to the electrokinetic force applied to the fluids, Eq.~\eqref{eq:electric_fluid_force}. We study the dependency of the contact angle $\theta$ on the strength of the applied external field $E_z^e \in (0,0.2) \, m \Delta x / (\Delta t)^2 e$, while varying the Gibbs transfer energy $\Delta \tilde \mu \in (-4,0)$, and bulk drop ion concentrations $\bar n_w = \left<n(\vect{r})\right> \in (10^{-4},10^{-2})/\Delta x^3$. This range of ionic densities sets the ratio of the drop radius with the Debye length $\xi = r/\lambda_D = (\varepsilon k_B T/\bar n_w e^2)^{1/2}/r \in (0.03,1)$, where we set $\varepsilon_r = \varepsilon / \varepsilon_0 = 80$, with $\varepsilon_0$ the permittivity of vacuum. Taking the lattice length scale as a nanometer, $\Delta x = 10^{-9}\,\text{m}$, salt concentrations are dilute, i.e. in the $(10^{-4}, 10^{-2})\,\text{mol}/\text{l}$ range~\cite{chevalliot:2012,garbow:2004}. Considering the setup at room temperature $T = 298.15K$, and the average fluid density as that of water $\rho = 1000\,\text{kg}/\text{m}^3$, the time unit $\Delta t = \sqrt{\rho \Delta x^5/k_B T} = 1.55\times10^{-11}\,\text{s}$, and thus the total time of the simulations, around a million timesteps, is in the microsecond scale. Considering the charge unit as the electron charge, $e = 1.6\times10^{-19}\,\text{C}$, the applied voltage differences are in the $(0,12)\,\text{V}$ range, corresponding to values previously used in experiments~\cite{chevalliot:2012}.  The Bjerrum length $l_B = e^2/4 \pi \varepsilon k_B T \approx 0.72 \Delta x$, implying that ion-ion interactions, which in our model are ignored, would be relevant only at the sub-grid level. In the following we express quantities in lattice units, where $\Delta x = \Delta t = k_B T = e = 1$.

\begin{figure}
	\begin{center}
		\includegraphics[scale=1.0]{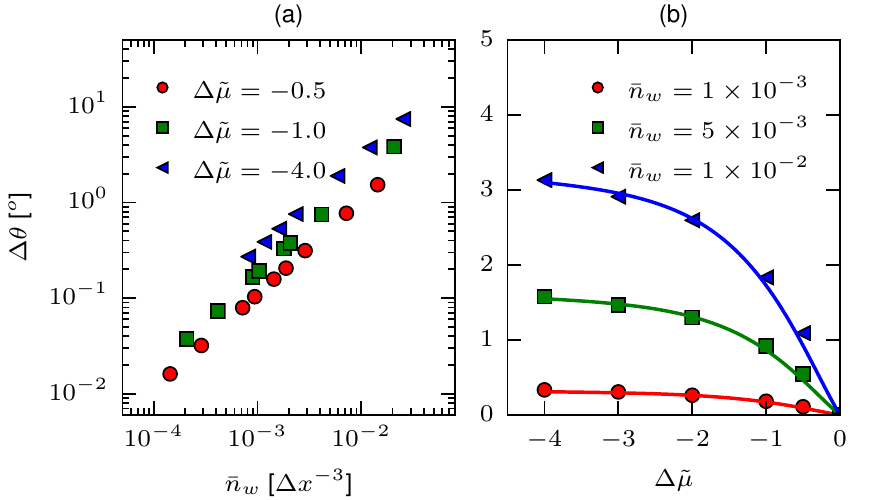}
	\end{center}
	\caption{(Color online) (a) Contact angle variations $\Delta \theta$ as a function of the salt concentration in the drop $\bar n_w$ (a), and Gibbs transfer energies $\Delta \tilde \mu$ (b), for no applied external field. Lines in (b) correspond to Eq.~\eqref{eq:yeahman}.}
	\label{fig:wetting_angle_e0}
\end{figure}

\textit{Results.---} We first study the dependency of the contact angle on the ion concentration and the Gibbs transfer energy for no applied electric field. The variation of $\theta$ with respect to the equilibrium contact angle with no salt, $\Delta \theta \equiv \theta - \theta_0$, is linear for two orders of magnitude in $\bar n_w$, as shown in Fig.~\ref{fig:wetting_angle_e0}(a). Moreover, for a given concentration of salt in water, a higher salt concentration in oil leads to smaller angle variations. This implies that the contact angle can be modified even for no applied external fields by varying the concentration of salt in the secondary phase.

The observed increase in contact angle can be explained mechanistically by deriving the variations of the electrokinetic force density $\vect{F}^e$ with the parameters $\Delta \tilde \mu$ and $\bar n_w$. Both quantities can be related via Young's law by adding the electrokinetic force to the force balance at the TCP, such that 
\begin{equation}
	\label{eq:deltatheta}
	\cos(\theta_0 + \Delta \theta) - \cos(\theta_0) = \mathcal{F}^e_y/r\gamma_{ow},
\end{equation}
with $\mathcal{F}^e_y \equiv \int F_y^e(y) \text{d}y$, the electrokinetic force in the $\hat y$ direction integrated through the drop interface. For small angle variations this results in $\Delta \theta \approx -\mathcal{F}^e_y/\gamma_{ow}r\sin\theta_0$. When no electric fields are present only the osmotic and solvation terms appear in Eq.~\eqref{eq:electric_fluid_force}, which together form a perfect gradient and therefore can be directly integrated. The total force is therefore invariant to the specific interface shape, and depends only on the bulk values of the ion and density concentrations for each species, such that 
$\mathcal{F}^e_y = - k_B T (n_o - n_w + n_o g \rho^w_o - n_w g \rho^w_w)$. As in equilibrium $\nu = \exp (-\Delta \tilde \mu)$~\cite{onuki:2006}, we obtain 
\begin{equation}
	\label{eq:yeahman}
	\Delta \theta = -\frac{k_B T n_w}{r \gamma_{wo}\sin\theta_0}
		\left[ 
 		  \exp(\Delta \tilde \mu) (1 + \Delta \tilde \mu \tilde \rho_o^a)
		- (1 + \Delta \tilde \mu \tilde \rho_w^a)
		\right].
\end{equation}

As expected, in the particular case of equally conducting fluids $\Delta \tilde \mu = 0$, $\Delta \theta = 0$. Furthermore, for a given $\Delta \tilde \mu$, $\Delta \theta \propto n_w$, as observed in Fig~\ref{fig:wetting_angle_e0}(a). As $\Delta \tilde \mu$ is decreased (increasing $|\Delta \tilde \mu|$), the exponential terms vanish and $\Delta \theta$ converges to a linear dependency on $\Delta \tilde \mu$. Fig.~\ref{fig:wetting_angle_e0}(b) shows that there is good agreement between Eq.~\eqref{eq:yeahman} and the measured variations of the contact angle. We also observe the same qualitative features and quantitative agreement with Eq.~\eqref{eq:yeahman} for hydrophobic surfaces, $\theta_0 = 135^o$.


\begin{figure}
	\begin{center}
		\includegraphics[scale=1.0]{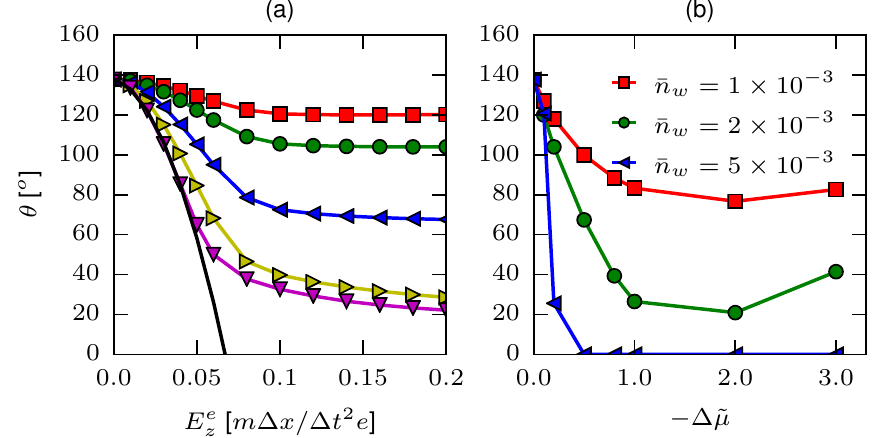}
	\end{center}
	\caption{(Color online) (a) Contact angle $\theta$ as a function of the applied external electric field $E_z^e$, for different Gibbs transfer energies $\Delta \tilde \mu=-0.1$ ($\square$), $\Delta \tilde \mu=-0.2$ ($\circ$), $\Delta \tilde \mu=-0.5$ ($\triangleleft$), $\Delta \tilde \mu=-1.0$ ($\triangleright$) and $\Delta \tilde \mu=-2.0$ ($\triangledown$) and drop ion concentration $\bar n_w = 2\times10^{-3}$. A quadratic dependence $\theta \propto (E_z^e)^2$ is shown in black. (b) Minimum contact angle attainable $\theta^*$ as a function of $\Delta \tilde \mu$ and different drop ion concentrations $\bar n_w$.}
	\label{fig:wetting_angle}
\end{figure}

Next, we consider the electrowetting setup in an external electric field. Here, the electric force deforms the drop such that it reduces its height and increases its width, as shown in Fig.~\ref{fig:drops}. As a consequence the contact angle $\theta$ decreases with increasing applied electric field $E_z^e$. For a given salt concentration in the drop, the extent of the deformation with the applied electric field, quantified by $\theta(E_z)$, increases with the Gibbs transfer energy $\Delta \tilde \mu$, as shown in Fig.~\ref{fig:wetting_angle}(a). Again, we observe that higher bulk ion concentration ratios $\nu$ lead to larger drop deformations.

In all cases studied, $\Delta \theta \propto (E_z^e)^2$ for $E^e_z \ll 1$, as is shown in Fig.~\ref{fig:wetting_angle}(a). Contrary to the no-field cases, it is not straightforward to derive a relation between the parameters and $\Delta \theta$, as given by Eq.~\eqref{eq:deltatheta}. Solutions of the full Poisson-Nernst-Planck system are complex even for a single fluid in the steady state case~\cite{golovnev2010steady}. Nevertheless, the quadratic decrease of $\theta$ can be traced to the formation of a Debye layer over the substrate, as ions of opposite charges diffuse in opposite directions with fluxes $D \beta q^\pm E_z$. For small ion concentrations such that $E_z^e \gg E_z^i$, then $q \propto E^e_z$. Together with the fact that $\phi_y^i \propto q$, the electric force at the TCP normal to the interface $q \vect{E} \cdot \hat y = q E_y^i \propto (E_z^e)^2$.


Contact angle saturation can be directly observed in Fig.~\ref{fig:wetting_angle}(a). Moreover, we find that the convergent contact angle $\theta^*$ decreases with increasing $\Delta \tilde \mu$, even to the point of complete wetting, as shown in Fig.~\ref{fig:wetting_angle}(b). That is, increasing the Gibbs transfer energy leads to the suppression of CAS. For the smallest (most negative) $\Delta \tilde \mu$ studied, we observe a change in behaviour of $\theta^*$, which we relate to finite size effects. The dominant term in this regard is the electrostatic repulsion experienced by the charges concentrated at the TCP, interacting through the periodic boundaries, which reduces the amount of deformation and therefore $\theta$. The repulsion increases with $\Delta \tilde \mu$ as it is proportional to the total charge in the drop. This implies that large drop deformations are here being underestimated, and thus the effect of $\Delta \tilde \mu$ can be expected to be even stronger in infinite systems.

\begin{figure}
	\begin{center}
		\includegraphics[scale=1.0]{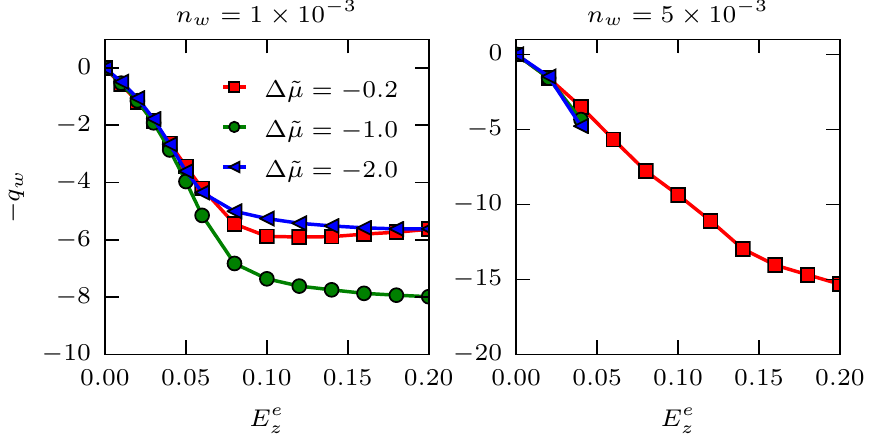}
	\end{center}
	\caption{(Color online) Variations of the total charge in the drop with respect to the applied field, $\partial q_w/\partial E_z^e$, as a function of $E^e_z$, for the indicated Gibbs transfer energies $\Delta \tilde \mu$, at $\bar n_w = 2\times10^{-3}$ (a) and $\bar n_w = 5\times10^{-3}$ (b).}
	\label{fig:ions_in_drop}
\end{figure}

Our observations indicate that CAS is a consequence of the depletion of positive ions in the drop, which leads to a convergence of the total charge. As the electric field is increased, positive ions are driven out of the drop (while negative ions accumulate in a Debye layer). Eventually, all positive ions can be expelled from the drop, leading to a convergence of the total drop charge $q_w$. Fig.~\ref{fig:ions_in_drop} shows the variations of $q_w$ with the applied field $E_z^e$ for hydrophobic substrates ($\theta_0 = 135^o$), which shows a strong initial decrease and then a sudden change in behaviour, to relatively small increases or decreases for low and high $\Delta \tilde \mu$, respectively. The point at which the change in behaviour takes place coincides with the change in $\Delta \theta(E_z^e)$. CAS is harder to observe for high ion concentrations as the drop completely wets the surface before the electric field is strong enough to remove all negative ions from the drop, as shown in Fig~\ref{fig:ions_in_drop}(b). All these observations also hold for $\theta_0 = 90^o$ (data not shown).

In the context of previously proposed explanations for CAS, our results are surprising taking into account the relative simplicity of our model. We have taken the substrate to be a perfect insulator, implying no transfer of ions between substrate and fluids, therefore discarding the proposed methods of dielectric discharging or breakdown~\cite{papathanasiou:2008,drygiannakis:2009}. Furthermore, in our simulations CAS cannot be the effect of the surface tension depending on the electric field or ion concentration because, as previously mentioned, in our model the intrinsic surface tensions are solely determined by fluid--fluid and fluid--substrate interaction parameters. We believe an interpretation in terms of effective surface tensions would be out of place here, as we have direct access to the forces involved and therefore a purely mechanistic approach is straightforward. Moreover, we observe no ejection of microdroplets near the TCP, as well as no Taylor cones~\cite{vallet:1999,park:2011,chevalliot2012experimental}. Interface instabilities might be frustrated by the resolution of the simulations and the dimensionality of the setup, although as mentioned, the interface thickness is not seen to vary significantly.



The general picture of CAS being a consequence of ion transfer is in agreement with recent studies~\cite{liu:2012,chen:2014,yamamoto:2015}. Molecular dynamics simulations of an electrowetting over a dielectric setup revealed that saturation is an effect of molecules being pulled out of the drop, which reduce the electric force at the TCP~\cite{liu:2012}. We have shown here that the molecular binding of the ions can be effectively modeled via a solvation potential, resulting in the same qualitative features. A theoretical derivation based on the Poisson-Boltzmann theory also shows that ionization of the oil phase leads to CAS~\cite{yamamoto:2015}. Our study represents a step forward in this direction, having solved the full set of electrokinetic equations considering a finite solvation potential.

\textit{Conclusion.---} We have shown that electrowetting of electrolyte suspensions can be controlled via selective solvation. Higher Gibbs transfer energies lead to stronger drop deformations, as the ion gradients and therefore the forces increase at the triple contact point. We have observed contact angle saturation to be a product of ion depletion from the drop, which leads to a convergent electric force at the triple contact points. The possibility to tune the convergent contact angle and even completely suppress contact angle saturation with the Gibbs transfer energy is of practical significance, especially in aqueous organic solutions where the Gibbs transfer energy can be modified by addition of solutes. In more general terms, we have demonstrated the feasibility to study electrowetting at nanoscales using a mesoscopic model, which can serve as the basis for future research which includes more resolved substrate-fluid interactions or differences in dielectric permittivities, as also, to study the dynamics of the deformations.

\textit{Acknowledgments.---} We thank the J\"ulich Supercomputing Centre and the High Performance Computing Centre Stuttgart for the technical support and the allocated CPU time.

\end{document}